\documentclass[conference]{IEEEtran}

\usepackage[T1]{fontenc}
\usepackage[scaled=0.85]{beramono}
\usepackage{newtxmath}

\usepackage{graphicx}

\usepackage{xcolor}

\usepackage[nobreak]{cite}
\usepackage{tablefootnote}
\usepackage{comment}
\usepackage{amsmath}
\usepackage{ascmac}
\usepackage{varwidth}
\usepackage{xspace}
\usepackage{xurl}
\usepackage{hyperref}

\makeatletter\let\MYcaption\@makecaption\makeatother
\usepackage[subrefformat=parens,font=footnotesize]{subcaption}
\makeatletter\let\@makecaption\MYcaption\makeatother

\usepackage{framed}
\newcommand{\Conclusion}[1]{\begin{framed}\noindent #1\end{framed}}

\usepackage{xstring}
\def\ModSplit#1{\IfSubStr{#1}{|}{\StrBefore{#1}{|}[\before]\StrBehind{#1}{|}[\behind]\ModItem{\before}\-\ModSplit{\behind}}{\ModItem{#1}}}
\def\ModItem#1{\texttt{#1}}

\def\Sym#1{\texttt{#1}}
\def\SymS#1{%
  \StrLen{#1}[\length]%
  \SymSImpl{#1}{\length}%
}
\def\SymSImpl#1#2{%
  \begingroup%
  \edef\W{#1}%
  \edef\B{#2}%
  \edef\A{\the\numexpr\B-1\relax}%
  \ifnum\B>1%
    \StrChar{\W}{\A}[\charA]%
    \StrChar{\W}{\B}[\charB]%
    \IfSubStr{abcdefghijklmnopqrstuvwxyz0123456789}{\charA}{%
      \IfSubStr{ABCDEFGHIJKLMNOPQRSTUVWXYZ}{\charB}{%
        \StrLeft{\W}{\A}[\beforeA]%
        \StrMid{\W}{\B}{10000}[\afterB]%
        \SymS{\beforeA}\-\Sym{\afterB}%
      }{\SymSImpl{\W}{\A}}%
    }{\SymSImpl{\W}{\A}}%
  \else%
    \Sym{\W}%
  \fi%
  \endgroup%
}
\newcommand{\Ident}[1]{\Sym{#1}}
\newcommand{\IdentM}[1]{\ModSplit{#1}}
\newcommand{\IdentS}[1]{\SymS{#1}}

\newcommand{\repo}[1]{\textit{#1}}
\newcommand{\app}[1]{\textit{#1}}

\def\field{$\mathit{field}$\xspace}
\def\method{$\mathit{method}$\xspace}
\def\enclosingClass{$\mathit{enclosingClass}$\xspace}
\def\parameter{$\mathit{parameter}$\xspace}
\def\enclosingMethod{$\mathit{enclosingMethod}$\xspace}
\def\pass{$\mathit{pass}$\xspace}

\def\parent{$\mathit{parent}$\xspace}
\def\ancestor{$\mathit{ancestor}$\xspace}

\def\assignmentEquation{$\mathit{assignmentEquation}$\xspace}
\def\argument{$\mathit{argument}$\xspace}

\def\type{$\mathit{type}$\xspace}

\def\siblingMembers{$\mathit{siblingMembers}$\xspace}
\def\parameterOverload{$\mathit{parameterOverload}$\xspace}

\def\Insert{$\mathit{insert}$}
\def\delete{$\mathit{delete}$}
\def\replace{$\mathit{replace}$}
\def\order{$\mathit{order}$}
\def\format{$\mathit{format}$}
\def\abbreviation{$\mathit{format}_a$}
\def\conjugation{$\mathit{format}_c$}
\def\plural{$\mathit{format}_p$}

\usepackage{cleveref}
\crefname{figure}{Fig.}{Figs.}
\crefname{table}{Table}{Tables}
\crefname{section}{Section}{Sections}
\Crefname{figure}{Figure}{Figures}

\begin{document}

\title{RENAS: Prioritizing Co-Renaming\\Opportunities of Identifiers}

\author{\IEEEauthorblockN{Naoki Doi}
\IEEEauthorblockA{\textit{School of Computing}\\
\textit{Tokyo Institute of Technology}\\
Tokyo, Japan \\
doi@se.c.titech.ac.jp}
\and
\IEEEauthorblockN{Yuki Osumi}
\IEEEauthorblockA{\textit{School of Computing}\\
\textit{Tokyo Institute of Technology}\\
Tokyo, Japan\\
osumi@se.c.titech.ac.jp}
\and
\IEEEauthorblockN{Shinpei Hayashi}
\IEEEauthorblockA{\textit{School of Computing}\\
\textit{Tokyo Institute of Technology}\\
Tokyo, Japan\\
hayashi@c.titech.ac.jp}
}

\maketitle

\pagestyle{plain}
\thispagestyle{plain}

\begin{abstract}
Renaming identifiers in source code is a common refactoring task in software development.
When renaming an identifier, other identifiers containing words with the same naming intention related to the renaming should be renamed simultaneously.
However, identifying these related identifiers can be challenging.
This study introduces a technique called RENAS, which identifies and recommends related identifiers that should be renamed simultaneously in Java applications.
RENAS determines priority scores for renaming candidates based on the relationships and similarities among identifiers.
Since identifiers that have a relationship and/or have similar vocabulary in the source code are often renamed together, their priority scores are determined based on these factors.
Identifiers with higher priority are recommended to be renamed together.
Through an evaluation involving real renaming instances extracted from change histories and validated manually, RENAS demonstrated an improvement in the F1-measure by more than 0.11 compared with existing renaming recommendation approaches.
\end{abstract}

\section{Introduction}\label{s:introduction}
Renaming identifiers is a crucial activity in software development, as identifiers constitute approximately 70\% of the source code \cite{Deissenbock-SQJ2006}, playing a pivotal role in developers' understanding of programs \cite{Schankin-ICPC2018}.
Inconsistent or inappropriate identifiers can significantly hinder developers' understanding, leading them to frequently modify these identifiers to more suitable names throughout the software development process \cite{Deissenbock-SQJ2006}.

However, renaming is not an easy task \cite{Arnaoudova-TSE2014}.
When a developer decides to rename an identifier, other associated identifiers that share the same concept are often necessary to be renamed as well.
This process can be time-consuming, as these identifiers may be spread across multiple files, and variations in word forms, such as inflections, must also be considered.
Failing to consistently rename an identifier can result in compromised naming consistency and a decrease in the internal quality of the software.

To reduce the effort, methods that identify and recommend groups of identifiers to be renamed together have been investigated\cite{Liu-TSE2015, Zhang-TOSEM2023}.
For example, Liu et al.\cite{Liu-TSE2015} specified identifiers related to the one being renamed and considered them as potential candidates for renaming, followed by the recommendation of identifiers to be renamed.
This process continues with identifiers associated with those recommended for renaming, with recommendations being made until the process fails.

Merely listing identifiers that can undergo the same renaming operation is insufficient for capturing all the identifiers that necessitate renaming (see \cref{c:motivation} for details).
Therefore, it is essential to include identifiers related to identifiers to which the renaming operation cannot be applied as candidates and to exclude identifiers that should not be renamed from the list of candidates.

This study proposes the renaming recommendation approach, referred to as RENAS, which determines priority scores for identifiers based on their relationships to the renamed identifiers.
By including identifiers that can be traced back through relationships in the recommendation scope, the approach identifies necessary renamings while assigning lower priority to identifiers named with different intentions.
We evaluate RENAS using real-world examples of renaming.
We show that RENAS is more accurate than existing approaches and that both the relationship and similarity should be taken into account for priority scores.

The main contributions of this study are as follows:
\begin{itemize}
  \item An approach for recommending identifiers with degrees of priority.
  \item A dataset of manually validated real-world co-renaming examples.
  \item A comparative evaluation of renaming recommendation approaches across 13 open-source software projects, including the application to the manually validated dataset.
\end{itemize}
The replication package, which includes the experimental dataset and the RENAS tool, is publicly available\cite{dataset,tool}.

The remainder of this paper is organized as follows.
\Cref{c:motivation} identifies problems of existing approaches using actual renamings.
\Cref{c:technique} explains the generation of renaming recommendations.
\Cref{c:evaluation} evaluates RENAS.
\Cref{c:relatedwork} reviews previous studies related to identifiers and their renaming practices.
\Cref{c:conclusion} concludes the paper and explores potential future research directions.

\section{Motivation}\label{c:motivation}

\begin{figure}[tb]\centering
    \begin{minipage}[tb]{0.5\textwidth}\centering
        \includegraphics[width=\linewidth]{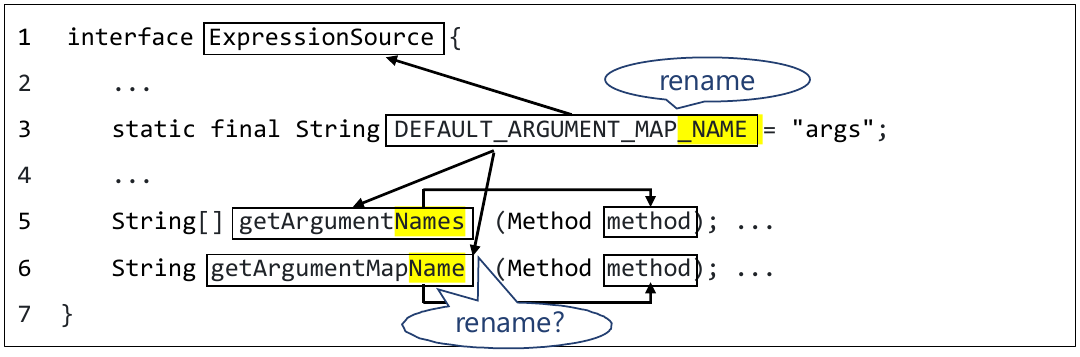}
        \subcaption{Before renaming.}
        \label{fig:motivation-rename1-before}
    \end{minipage}\vspace{0.5em}
    \begin{minipage}[tb]{0.5\textwidth}\centering
        \includegraphics[width=\linewidth]{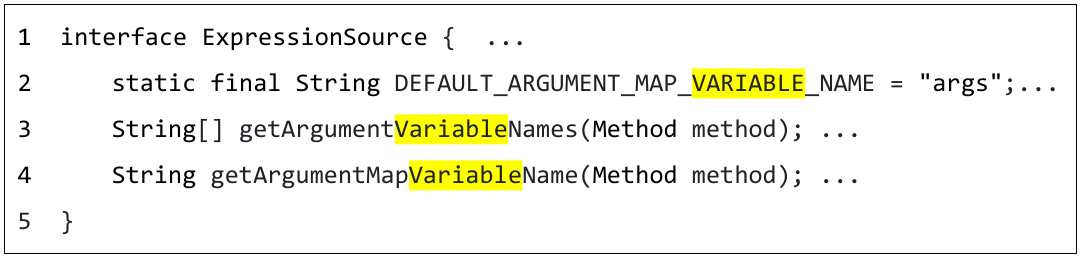}
        \subcaption{After renaming.}
        \label{fig:motivation-rename1-after}
    \end{minipage}
    \caption{Renaming in \repo{spring-integration}.}\label{fig:motivation-rename1}
\end{figure}

Renamings are a common practice in software development, although they are not always straightforward \cite{Arnaoudova-TSE2014}.
When one identifier is renamed, other identifiers named with the same intention should also be co-renamed.
An example of renaming in \repo{spring-integration}\footnote{\url{https://github.com/spring-projects/spring-integration/commit/5ebafd8}} is shown in \cref{fig:motivation-rename1}.
The source code before and after renaming are shown in \cref{fig:motivation-rename1-before,fig:motivation-rename1-after}, respectively.
The changes made before and after renaming are highlighted, showcasing the insertion of the term \Ident{variable} before the term \Ident{name}.
The field name \IdentM{DEFAULT\_|ARGUMENT\_|MAP\_|NAME}, method name \IdentS{getArgumentNames}, and \IdentS{getArgumentMapName} contain the word \Ident{name}, and because they indicate the same concept, the developer insert \Ident{variable} for them as well.
This practice of co-renaming is quite common, as reported by Saika et al.\cite{Saika-IWESEP2014}, who found that renaming was the most frequent refactoring conducted simultaneously.
Finding identifiers that should be co-renamed can be a time-consuming task, and overlooking such changes may lead to inconsistencies in the code.

To reduce the effort, existing approaches have been developed to recommend identifiers that should be co-renamed based on their relationships with the renamed identifier. 
For example, Liu et al.\cite{Liu-TSE2015} defined relationships between identifiers and extracted candidate identifiers to which the same renaming operation can be applied for renaming based on the defined relationships with the renamed identifier, prioritizing identifiers with high similarity to the renamed identifier.

We present an example of an application of a straightforward relationship-based approach.
The initial step for the source code shown in \cref{fig:motivation-rename1-before} involves normalizing \Ident{names} to \Ident{name} by eliminating the word form from the renaming.
For example, when a developer renames \IdentM{DEFAULT\_|ARGUMENT\_|MAP\_|NAME} to \IdentM{DEFAULT\_|ARGUMENT\_|MAP\_|VARIABLE\_|NAME}, the renaming operation ``insert \Ident{variable} before \Ident{name}'' is obtained.
From the relationships indicated as arrows in \cref{fig:motivation-rename1-before}, after eliminating the word form, the following candidate identifiers are obtained: \IdentS{ExpressionSource}, \IdentS{getArgumentName}, and \IdentS{getArgumentMapName}.
Subsequently, we check whether the renaming operation can be applied to these identifiers.
The identifiers \IdentS{getArgumentName} and \IdentS{getArgumentMapName} are found to be applicable, and unexplored identifiers associated with them are added to the list of candidates.
The newly identified candidates are two \Ident{method} parameters in Lines 5 and 6.
However, because the renaming operation cannot be applied to them, the recommendation process is halted, resulting in the final outcome of (\IdentS{getArgumentNames}, \IdentS{getArgumentMapName}).

\begin{figure}[tb]\centering
    \begin{minipage}[tb]{0.5\textwidth}\centering
        \includegraphics[width=8.6cm]{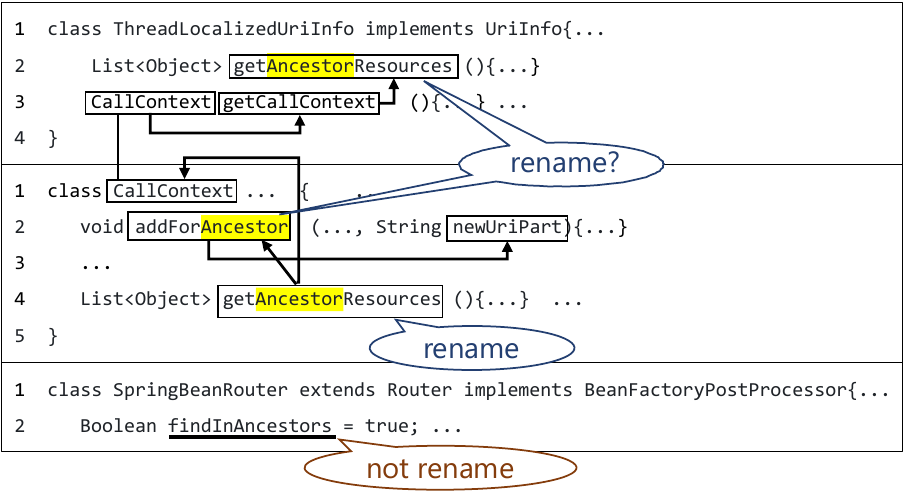}
        \subcaption{Before renaming.}
        \label{fig:motivation-example-rename2-before}
    \end{minipage}\vspace{0.5em}
    \begin{minipage}[tb]{0.5\textwidth}\centering
        \includegraphics[width=8.6cm]{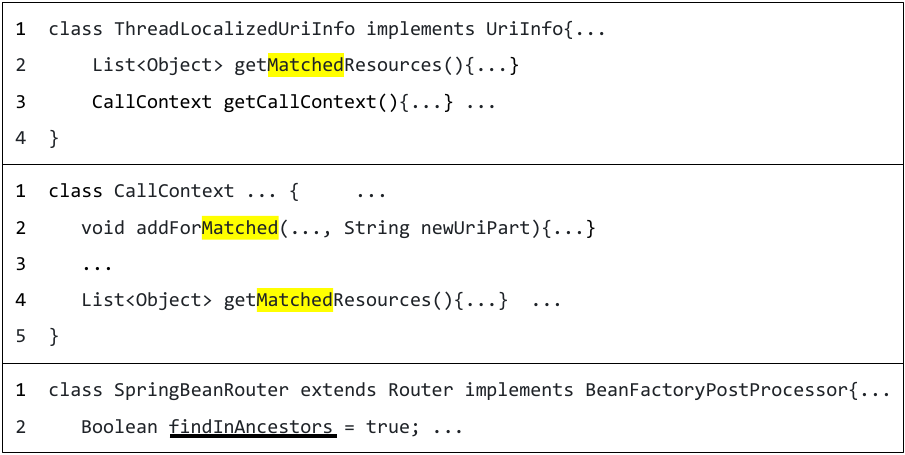}
        \subcaption{After renaming.}
        \label{fig:motivation-example-rename2-after}
    \end{minipage}
    \caption{Renaming in \repo{restlet-framework-java}.}
    \label{fig:motivation-example-rename2}
\end{figure}

\begin{figure*}[tb]\centering
  \includegraphics[width=10cm]{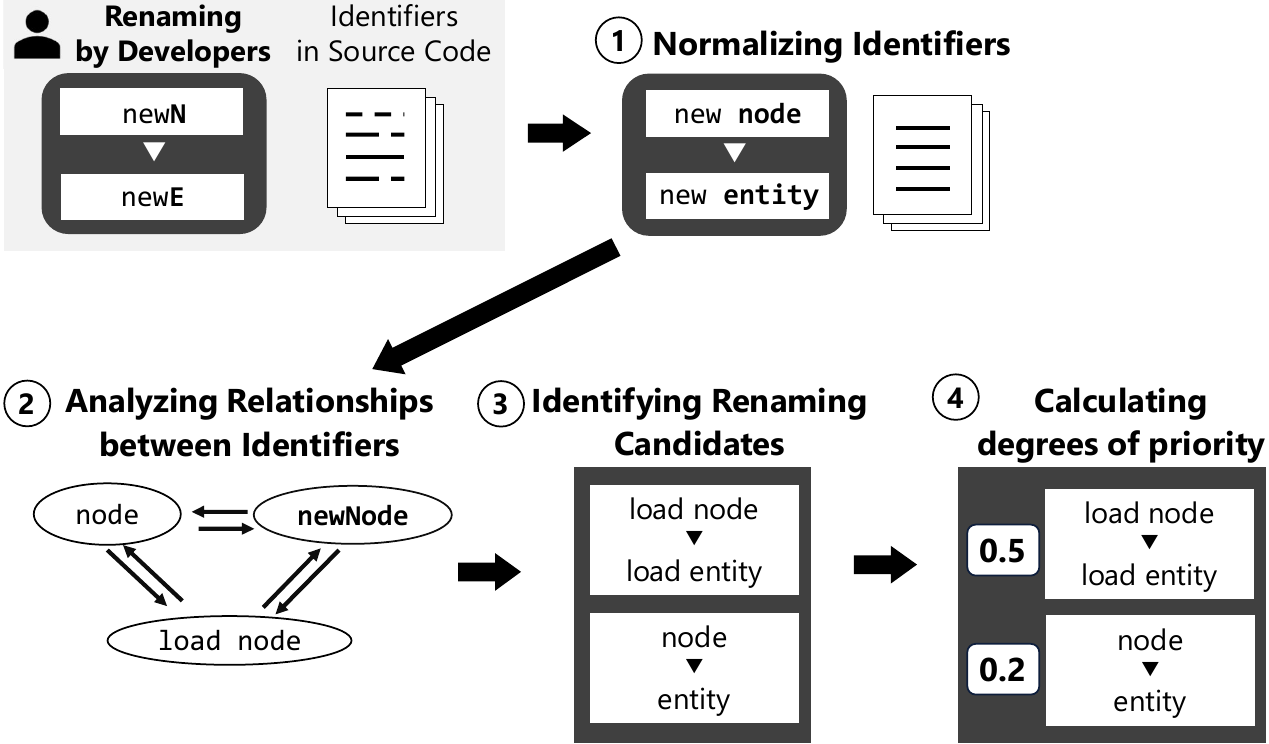}
  \caption{Overview of RENAS.}
  \label{fig:overview}
\end{figure*}

Existing approaches cannot enumerate all the identifiers that require renaming.
The actual example of renaming performed in \repo{restlet-framework-java}\footnote{\url{https://github.com/restlet/restlet-framework-java/commit/2938644}} is shown in \cref{fig:motivation-example-rename2}.
The source code before and after renaming are shown in \cref{fig:motivation-example-rename2-before,fig:motivation-example-rename2-after} respectively, with the renamed identifiers highlighted.
These renamings altered the concept of \Ident{ancestor} to \Ident{matched}.

For example, consider renaming the method \IdentS{getAncestorResources} of the class \IdentS{CallContext} to \IdentS{getMatchedResources}.
By examining the relationships indicated as arrows in \cref{fig:motivation-example-rename2}, the potential identifiers are \IdentS{CallContext} and \IdentS{addForAncestor}.
Additionally, because the renaming operation can be applied to \IdentS{addForAncestor}, the identifier \IdentS{newUriPart}, which is related to it, is also a candidate.
However, the recommendation ends here because the renaming operation cannot be applied.
The result consists only of \IdentS{addForAncestor}, and \IdentS{getAncestorResources} of class \IdentS{ThreadLocalizedUriInfo} is not recommended.

To find all identifiers requiring renaming, candidates should be explored also from identifiers related to the identifiers to which the renaming operation cannot be applied. 
By analyzing the relationships in \cref{fig:motivation-example-rename2-before}, three relationships can be traced from \IdentS{getAncestorResources} of \IdentS{CallContext} to \IdentS{getAncestorResources} of \IdentS{ThreadLocalizedUriInfo}.
Consequently, by tracing relationships from the identifiers to which the renaming operation cannot be applied, candidates can include \IdentS{getAncestorInfos} of \IdentS{ThreadLocalizedUriInfo}.

However, excessive relationship tracing can result in inappropriate identifier recommendations.
In \cref{fig:motivation-example-rename2-before}, the \IdentS{findInAncestors} of the class \IdentS{SpringBeanRouter} was not renamed.
The \Ident{Ancestors} in \IdentS{findInAncestors} was not renamed because it represents a different concept from the renamed \Ident{Ancestor} in \IdentS{getAncestorResources}.
The \IdentS{findInAncestor} can be reached by tracing the relationship from \IdentS{getAncestorResources} more than ten times.
Therefore, if unlimited transitions of relationships are permitted, the recommendation of \IdentS{findInAncestors} would occur.

In this study, we propose an approach called RENAS, which assigns priority to all identifiers that can be traced from the renamed identifier.
This approach utilizes this priority to recommend renamings.
The priority assigned is based on the similarity of identifiers and the number of traced relationships.
By tracing relationships regardless of the application of the renaming operation, we can recommend identifiers that existing approaches overlook.
Furthermore, by assigning priority to identifiers, those named with different intentions are given lower priority and are less likely to be recommended.

\section{Methodology}\label{c:technique}

\subsection{Overview}\label{s:overviewTechnique}

An overview of RENAS is presented in \cref{fig:overview}.
The input for this approach consists of Java source code and an identifier that has been renamed by developers.
The output includes a list of identifier candidates that can be renamed and their priority levels.
First, identifiers in the source code and the identifier renamed by developers are normalized, followed by the analysis of relationships between identifiers in the source code.
Next, the relationships are traced from the renamed identifier.
Subsequently, the renaming candidates are extracted to include identifiers to which the same renaming operation can be applied.
Finally, RENAS calculates the priority of the candidates and provides recommendations.
RENAS was designed to recommend renaming opportunities of identifiers for developers, rather than generating new names for identifiers.
RENAS can recommend all identifiers whose priority levels are equal to or greater than a specified threshold as a set, or they can opt to receive recommendations in a ranked order based on their priority levels.

\def\lowerit{\textit{lower}\xspace}
\def\Titleit{\textit{Title}\xspace}
\def\UPPERit{\textit{UPPER}\xspace}

\subsection{Normalizing Identifiers}\label{s:normalizeIdentifier}
We perform normalization by splitting identifiers, expanding abbreviations, and eliminating inflection in a specific order.
This process is performed on identifiers in the source code and the renamed identifier.

We split all identifier names into lowercase word sequences.
Identifier names named in \textit{snakeCase}, \textit{camelCase}, and \textit{PascalCase} can be divided into a sequence of words.
First, we set the delimiter characters to \texttt{\$}, \texttt{\_}, and numbers, and split identifier names accordingly.
Next, we divide words based on \lowerit, \Titleit, and \UPPERit word forms.
The \lowerit comprises all lowercase letters, \Titleit is a word with only the first letter capitalized, and \UPPERit comprises all uppercase letters.
Finally, all words are then unified into lowercase.

Abbreviations are expanded using KgExpander\cite{Jiang-FSE2019} with an extension.
KgExpander finds words before being abbreviated by tracing relationships between identifiers.
We added a new relationship to KgExpander (details provided in the following subsection).
When expanding an abbreviation, we maintain a record of the expansion, including the abbreviation, words before the abbreviation, and the source file containing the abbreviation.

\begin{figure*}[tb]\centering
    \includegraphics[width=12cm]{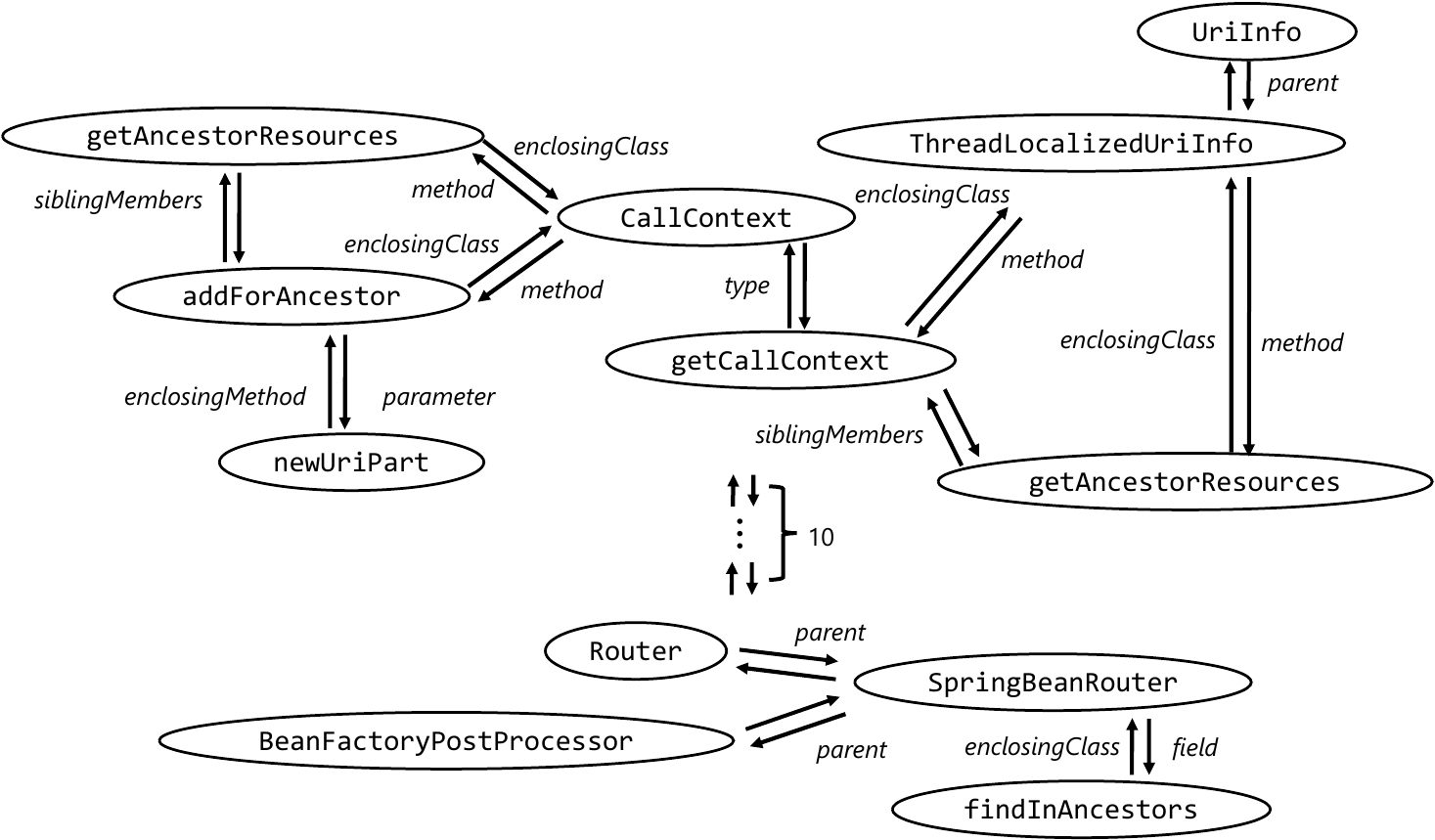}
    \caption{Relationship graph.}\label{fig:relationship-graph}
\end{figure*}

When an identifier name after renaming contains an abbreviation, four steps should be followed to find words before expanding the abbreviation.
\begin{enumerate}
    \item Searching from words in the identifier name before it was renamed.
    \item Searching from the expansion history of the same file\label{s:abbreviationProcess2}:
    Abbreviation candidates are obtained from the expansion history of the source file where renaming occurred.
    In the case of multiple candidates, the one with the most number of expansions is selected.
    Suppose that the number of expansions is the same.
    In that case, we select the one with the longest word length.
    \item Searching by the expansion history of the entire project:
    We select the one with the longest word length from the entire project's history.
    \item Using abbreviation dictionaries:
    We utilize the dictionaries provided by KgExpander.
    Note that we excluded commonly abbreviated words, such as PDF and UML, from these dictionaries.
\end{enumerate}

Finally, NLTK\footnote{\url{https://www.nltk.org/}}\cite{nltk} unifies the conjugated forms of verbs with the infinitive form, plural of nouns with singular, and comparative and superlative degrees of adjectives with the positive degrees.

\subsection{Analyzing Relationships between Identifiers}\label{s:relationIdentifier}

We analyze the following relationships to find identifiers to be renamed.
The extraction of relationships is performed at the project level by static analysis of all the source files of a Java project before the commit, excluding those defined in external libraries, using the ASTParser of Eclipse JDT\footnote{\url{https://projects.eclipse.org/projects/eclipse.jdt}}.
The relationships to be extracted are those used in KgExpander\cite{Jiang-FSE2019} (Inclusion, Inheritance, Assignment, and Typing) and Sibling used in RenameExpander\cite{Liu-TSE2015}.
\begin{itemize}
    \item Inclusion
    \begin{itemize}    
        \item \field: Relationship from a class $i_c$ to its field $i_f$ ($i_c \to i_f$)
        \item \method: Relationship from a class $i_c$ to its method $i_m$ ($i_c \to i_m$)
        \item \enclosingClass: Relationship from a field $i_f$ and a method $i_m$ to a class $i_c$ enclosing it ($i_f, i_m \to i_c$)
        \item \parameter: Relationship from a method $i_m$ to its parameter $i_p$ ($i_m \to i_p$)
        \item \enclosingMethod: Relationship from a parameter $i_p$ and a local variable $i_v$ to a method $i_m$ enclosing it ($i_p, i_v \to i_m$)
        \item \pass: Relationship from an argument (actual parameter) $i_a$ of a method $i_m$ to the method $i_m$ ($i_a \to i_m$)
    \end{itemize}
        
    \item Inheritance
    \begin{itemize} 
        \item \parent: Relationship between a class $i_c$ and its direct subclass $i_\mathit{ec}$ or its interface $i_\mathit{ic}$ ($i_c \leftrightarrow i_\mathit{ec}, i_\mathit{ic}$)
        \item \ancestor: Relationship between a class $i_{c}$ and its ancestor class (a subclass of its subclass or more) $i_\mathit{ac}$ ($i_{c} \leftrightarrow i_\mathit{ac}$)
    \end{itemize}
    
    \item Assignment
    \begin{itemize}
        \item \assignmentEquation: Relationship between a left side value $i_\mathit{left}$ and a right side value $i_{right}$ of the assignment statement, such as a field, parameter, variable, or invocation of a method ($i_\mathit{left}$ $\leftrightarrow$ $i_\mathit{right}$)
        \item \argument: Relationship between a parameter $i_p$ of a method $i_m$ and an argument (actual parameter) $i_a$ of the method $i_m$ ($i_p \leftrightarrow i_a$)
    \end{itemize}

    \item Typing
    \begin{itemize}
       \item \type: Relationship between an identifier $i$, such as a field, method, parameter, or variable and its type $i_t$ ($i \leftrightarrow i_t$)
    \end{itemize}
    
    \item Sibling
    \begin{itemize}
        \item \siblingMembers: Relationship between fields $i_f$ of a class $i_c$ and methods $i_m$ of a class $i_c$($i_f \leftrightarrow i_f, i_m \leftrightarrow i_m, i_f \leftrightarrow i_m$)
        \item \parameterOverload: Relationship between a parameter $i_{p1}$ of a method $i_{m1}$ and a parameter $i_{p2}$ of a method $i_{m2}$ that overloads $i_{m1}$ ($i_{p1} \leftrightarrow i_{p2}$)
    \end{itemize}
\end{itemize}

Notably, RenameExpander does not utilize Typing, and other relationships it employs are similar but not identical.
For example, Inheritance only considers parent classes in RenameExpander.
Additionally, we introduced a new relationship \parameterOverload to the existing relationships used by RenameExpander.
Among the methods defined in the class, the overloaded methods $i_{m1}$ and $i_{m2}$ exhibit functional similarities, and the identifier vocabulary used is largely identical.
Consequently, we defined this relationship as a shortcut to the chain of (\enclosingMethod, \siblingMembers, \parameter) to prioritize recommendations.

To illustrate these relationships, \cref{fig:relationship-graph} displays the relationships between identifiers in \cref{fig:motivation-example-rename2-before}.
The upper and lower graphs represent different segments of the relationship graph, with identifiers in the lower graph being traced back ten times from those in the upper graph.
For example, the path from \IdentS{addForAncestor} to \IdentS{getCallContext} can be reached by following two relationships (\enclosingClass, \type).

\subsection{Identifying Renaming Candidates}\label{s:recommendName}
We extract renaming operations from the developer's renaming. 
When an identifier is renamed, we find other identifiers that are reachable via relationships from it and to which one of the extracted renaming operations is applicable.
We consider them as potential candidates for renaming.

We define the following five types of renaming operations.
\begin{itemize}
    \item \Insert(inserted word sequence): Insert word sequence. \\
        \;\Ident{reference}$\to$\Ident{\underline{word}Reference}:
 \Insert([\Ident{word}])
    \item \delete(deleted word sequence): Delete word sequence. \\
    \;\Ident{\underline{word}Reference}$\to$\Ident{reference}: \delete([\Ident{word}])
    \item \replace(word sequence before replacement, word sequence after replacement): Replace word sequence. \\
        \;\Ident{\underline{term}Ref} $\to$ \Ident{\underline{word}Ref}: 
 \replace([\Ident{term}], [\Ident{word}])
    \item \order(word sequence before reordering, word sequence after reordering): Reorder word sequence. \\
        \;\Ident{refWord}$\to$\Ident{wordRef}: 
 \order([\Ident{ref}, \Ident{word}], [\Ident{word}, \Ident{ref}])
    \item \format(word before format change, word after format change): Change word format
    \begin{itemize}
        \item \plural: Change the plural or singular form of a word. 
            \; \Ident{word} $\to$ \Ident{word\underline{s}}:  \plural(\Ident{word}, \Ident{words})
        \item \abbreviation: Abbreviate word. \\
            \; \Ident{\underline{ref}erence} $\to$ \Ident{ref}: \abbreviation(\Ident{reference}, \Ident{ref})
        \item \conjugation: Change word conjugation. \\
            \; \Ident{refer} $\to$ \Ident{refer\underline{ing}}:  \conjugation(\Ident{refer}, \Ident{refering})
    \end{itemize}
\end{itemize}

When multiple renaming operations are extracted from a renaming, the renaming is considered as a candidate if at least one extracted operation is applicable.
However, certain renaming operations, such as customary changes, typo corrections, changes in the structure of identifier names, and expansion of abbreviations do not adhere to the relationships \cite{Zhang-TOSEM2023}.
Therefore, we exclude renaming operations such as \order, \abbreviation, or \conjugation\ from the recommendations.
The applicable conditions for each operation are as follows:
\begin{itemize}
    \item $\mathit{insert}$: The inserted word sequence matches either before or after.
    \item \delete: Deleted word sequence exists.
    \item \replace: Word sequence before replacement exists.
    \item \order: Two or more words are contained in a word sequence before reordering, and the order of appearance differs.
    \item \format: Word before format change exists.
\end{itemize}

\def\Op{\mathit{Op}}
\def\op{\mathit{op}}
\def\Name{\mathrm{name}}
We define the set of renaming refactorings as $R_{\Name}=\{r_0,r_1,\dots,r_{n-1}\}$, and $\Op(r)$ represents the renaming operations in the refactoring $r$.
Let $S_{\op}=\{r \in R_{\Name} \mid \op \in \Op(r)\}$ be the co-renamed set when the renaming operation is common.
If two or more renaming operations exist, the renaming belongs to more than one set.

\subsection{Calculating Priority of Renaming Candidates}\label{s:recommendRange}
We determine the priority of identifiers within the renaming candidates.
The higher the priority, the more the identifier should be renamed.

We calculate the priority based on the similarity and relationship of identifiers.
In \cref{ss:similarity}, we investigate the similarity between identifiers to highlight its significance in determining the priority for recommending co-renaming of identifiers.
In addition, we calculate the priority considering relationships because identifiers connected by relationships are often co-renamed \cite{Osumi-APSEC2022}.

\subsubsection{Calculating Priority Based on Similarity}\label{ss:similarity}

\def\Change{\mathrm{change}}
\def\Cand{\mathrm{cand}}
\def\Dice{\mathit{Dice}}
\def\Score{\mathit{Score}}
\def\Sim{\mathrm{sim}}
\def\ScoreSim{\Score_\Sim}
\def\ScoreRel{\Score_\mathrm{rel}}

The similarity between the renamed identifier $I_{\Change}$ and candidate identifier $I_{\Cand}$ (both normalized) is computed using the Dice coefficient, which is based on the similarity of constituent words.
This coefficient is preferred over other similarity indices to highlight shared vocabulary.
The similarity score is defined as follows:
\begin{align} 
    \ScoreSim(I_\Change, I_\Cand) &= \Dice(I_\Change, I_\Cand) \notag \\
        &= \frac{2 * |I_\Change \cap I_\Cand|}{|I_\Change|+|I_\Cand|}. \label{eq:similarity}
\end{align}

The co-renamed identifiers exhibited high $\ScoreSim$.
Using RefactoringMiner 2.0.2\cite{Tsantalis-TSE2022}, we generated co-renamed sets $S_{\op}$ from the renaming data.
The names of four projects used in this investigation, the number of co-renamed sets, and the number of renamed identifiers are shown in \cref{tab:4repository}.
To ensure a diverse range of projects for the primary evaluation, four projects were randomly selected from the dataset of Osumi et al.\cite{Osumi-APSEC2022}; coincidentally, two projects were small, whereas the remaining two were voluminous.
Although the number of projects may not be extensive, we believe they are highly effective in promoting the utilization of $\ScoreSim$ and the threshold settings surveyed in \cref{ss:hyperparameter}.

\begin{figure}[tb]\centering
    \includegraphics[width=6cm]{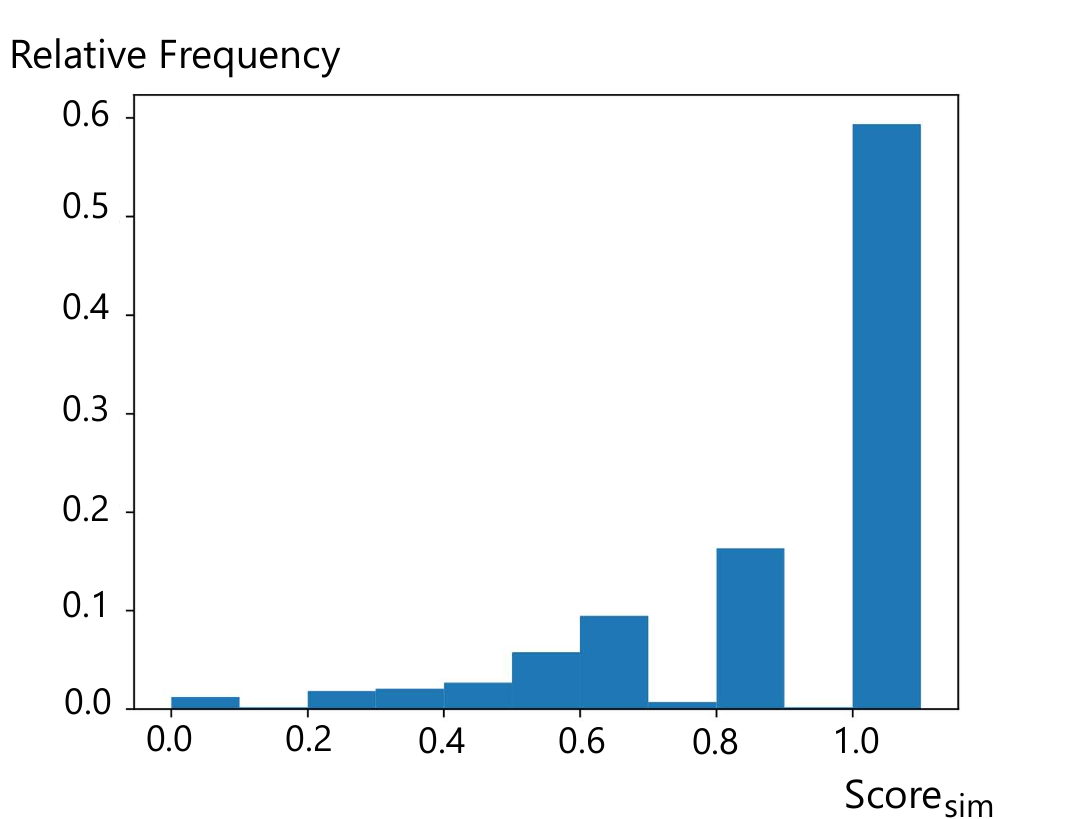}
    \caption{Distribution of $\ScoreSim$.}\label{fig:similarity_histgram}
\end{figure}

We calculated $\ScoreSim$ of identifier names prior to renaming for all combinations $\{(r_i,r_j) \mid r_i,r_j\in S_{\op} \land i < j\}$ within the set.
The result is shown in the histogram in \cref{fig:similarity_histgram}.
The vertical axis represents the relative frequency, whereas the horizontal axis represents $\ScoreSim$ across 11 classes $\{[x/10,(x+1)/10) \mid 0 \leq x \leq 10, x \text{ is an integer}\}$.
This figure illustrates a positive correlation between higher $\ScoreSim$ values and increased relative frequency, suggesting that co-renamed identifiers often exhibit high $\ScoreSim$.

\begin{table}[tb]\centering
    \caption{Four Projects Utilized in the Investigation}\label{tab:4repository}
    {\begin{tabular}{lcc} \hline
       Project & \# renames & \# co-renamed sets \\\hline
       \repo{cordova-plugin-local-notifications} & 40 & 19\\
       \repo{morphia} & 1,308 & 260\\
       \repo{spring-integration} & 3,649 & 995\\
       \repo{baasbox} & 36 & 13\\\hline
    \end{tabular}}
\end{table}

\subsubsection{Calculating Priority Based on Relationships of Identifiers}\label{ss:relation}

\def\Distance{\mathit{distance}}
\def\Cost{\mathit{Cost}}
\def\Rel{\mathit{Rel}}
\def\rel{\mathit{rel}}
To calculate the distance on the relationship graph between the renamed identifier and renaming candidate, we sum the cost weights within the relationship graph as follows:
\begin{equation} \label{eq:distance}
    \Distance = \sum_{\rel \in \Rel}\Cost(\rel)
\end{equation}
where $\Rel=\{\rel_1, \rel_2, \dots\}$ represents the set of relationships traced on the shortest path, and 
$\Cost$ represents the cost function based on the type of relationship, as shown in \cref{tab:relationship_priority}.
We established four distinct $\Cost$ function based on the frequency of co-renaming\cite{Osumi-APSEC2022}.
Furthermore, we categorized the relationships of \pass, \ancestor, and \parameterOverload, which were not included in the result by Osumi et al.\cite{Osumi-APSEC2022}, by employing the same categories as other similar relationships.

\begin{table}[tb]\centering
    \caption{Priority of Each Relationship}\label{tab:relationship_priority}
    \begin{tabular}{cp{7cm}} \hline
       $\Cost$ & Relationship  \\\hline
       1 & \assignmentEquation, \siblingMembers, \parameterOverload \\
       2 & \argument\\
       3 & \parameter, \enclosingMethod, \pass, \parent, \ancestor, \type \\
       4 & \field, \method, \enclosingClass \\\hline
    \end{tabular}
\end{table}

The priority based on the relationship is obtained as the inverse of $\Distance$, prioritizing the shortest distance:
\begin{equation} \label{eq:distance-priority}
    \ScoreRel = \frac{1}{\mathrm{distance}}.
\end{equation}

\subsubsection{Calculating Overall Priority}\label{ss:hyperparameter}

\begin{figure}[tb]
    \centering
    \begin{minipage}[tb]{0.47\linewidth}
        \centering
        \includegraphics[width=4cm]{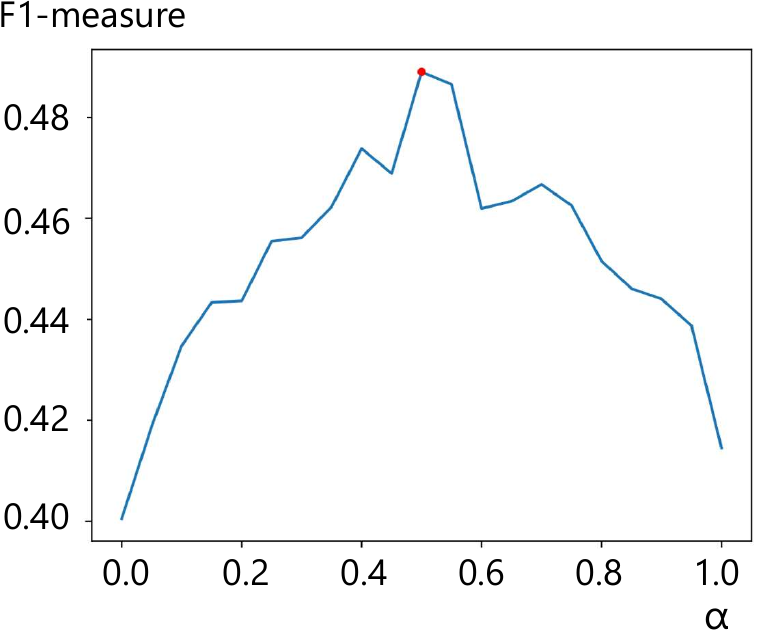}
        \caption{Maximum F1-measure for each $\alpha$ in \cref{eq:priority}.}
        \label{fig:alpha-graph}
    \end{minipage}
   \hfill
    \begin{minipage}[tb]{0.47\linewidth}
        \centering
        \includegraphics[width=4cm]{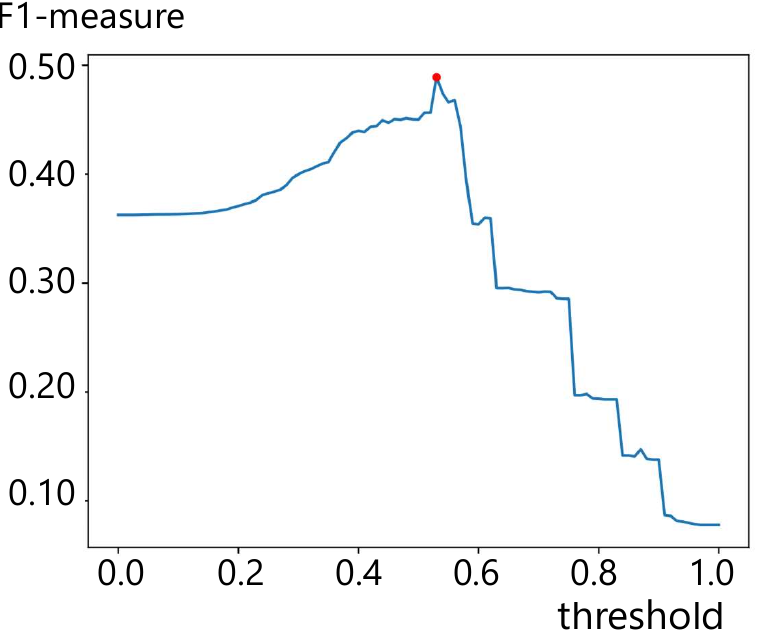}
        \caption{F1-measure at each threshold $\beta$ where $\alpha$ = 0.5.}
        \label{fig:alpha-F値}
    \end{minipage}
\end{figure}

The overall priority of an identifier, denoted as $\Score$, is determined by a linear combination of two priorities.
The priority can be calculated as follows:
\begin{equation} \label{eq:priority}
    \Score = \alpha \, \ScoreSim + (1 - \alpha) \, \ScoreRel.
\end{equation}
The parameter $\alpha$ represents the mixing ratio of $\ScoreRel$ and $\ScoreSim$.
The recommendation result can manifest as either a ranking based on priority or a set of identifiers with a priority equal to or greater than the threshold $\beta$.
The parameters $\alpha$ and $\beta$ were defined empirically.
Based on the approach in \cref{ss:similarity}, we generated co-renamed sets for the four projects in \cref{tab:4repository} and examined whether one renaming in the set could recommend other renamings.
Based on $\alpha \in \{ 0, 0.05, 0.10, \dots, 1 \}$ and $\beta \in \{ 0, 0.05, 0.10, \dots, 1 \}$, the results obtained by executing RENAS are shown in \cref{fig:alpha-graph,fig:alpha-F値}.
The largest F1-measure among all $\beta$ trials at a particular $\alpha$ is shown in \cref{fig:alpha-graph}.
From this graph, an optimal value of $\alpha = 0.5$ was selected, corresponding to an F1-measure of 0.489.
The F1-measure at each $\beta$ when $\alpha=0.5$ is shown in \cref{fig:alpha-F値}.
Based on the results, $\beta=0.53$ was adopted, taking the maximum value.

\subsubsection{Example}\label{ss:calculateCostExample}

We illustrate an example calculation when the method \IdentS{getAncestorResources} in the class \IdentS{CallContext} in \cref{fig:motivation-example-rename2-before} is renamed to \IdentS{getMatchedResources}.
Upon analyzing the relationship graph shown in \cref{fig:relationship-graph}, three identifiers exist that could apply the renaming operation:
\begin{itemize}
    \item \IdentS{addForAncestor}: (\siblingMembers),
    \item \IdentS{getAncestorResources}: (\enclosingClass, \type,\\\siblingMembers),
    \item \IdentS{findInAncestors}: (\enclosingClass, \type, $\ldots$).
\end{itemize}
The priority of each identifier is then calculated.
For example, when determining the priority of \IdentS{addForAncestor}, the $\Dice$ between [\Ident{add}, \Ident{for}, \Ident{ancestor}] and [\Ident{get}, \Ident{ancestor}, \Ident{resource}] is $1 \times 2 / (3 + 3) = 0.333$. 
Therefore, the priority $\ScoreSim$ calculated from the $\Dice$ is 0.333. 
The relationship \siblingMembers is traced from \IdentS{ancestorResources} to \IdentS{addForAncestor}.
Because the $\Cost$ of \siblingMembers is 1, the $\Distance$ calculated from the relationship is 1, resulting in a priority $\ScoreRel$ of 1, as per \cref{eq:distance-priority}.
Finally, the overall priority $\Score$ was determined to be $0.5 \times 0.333 + 0.5 \times 1 = 0.667$ as per \cref{eq:priority}.
Similarly, because $\ScoreSim$ of \IdentS{getAncestorResources} is $2 \times 3 / (3 + 3) = 1$, the $\Distance$ of the relationship is $4 + 3 + 1 = 8$ and $\ScoreRel$ is $1 / 8 = 0.125$.
Therefore, $\Score$ is $0.5 \times 1 + 0.5 \times 0.125 = 0.563$.
Furthermore, $\ScoreSim$ of \IdentS{findInAncestors} is $2 \times 1 / (3 + 3) = 0.333$ and the $\Distance$ of the relationship is $4 + 3 + \cdots \geq 8$ and $\ScoreRel \leq 1 / 8 = 0.125$. Therefore, $\Score \leq 0.5 \times 0.333 + 0.5 \times 0.125 = 0.229$.

In the case of renaming recommendation using the threshold $\beta$, the priority of \IdentS{addForAncestor} and \IdentS{getAncestorResources} exceeds the threshold of 0.53, resulting in two renaming recommendations being generated.

\section{Evaluation}\label{c:evaluation}

\def\RQone{What is the performance of RENAS compared with that of existing approaches?}
\def\RQtwo{How does the performance vary depending on how to prioritize?}

In this section, we evaluate RENAS by answering the two research questions.
\begin{itemize}
    \item RQ1: \RQone
    \item RQ2: \RQtwo
\end{itemize}
RQ1 aims to demonstrate that RENAS outperforms existing methods in terms of renaming recommendations.
In answering RQ1, we utilize the set of recommendations obtained using the threshold $\beta$.
RQ2 focuses on the validity of calculating priority based on identifier similarity and relationships.
RQ2 utilizes recommendation ranking based on priority.

\begin{figure}[tb]
    \centering
    \includegraphics[width=8.5cm]{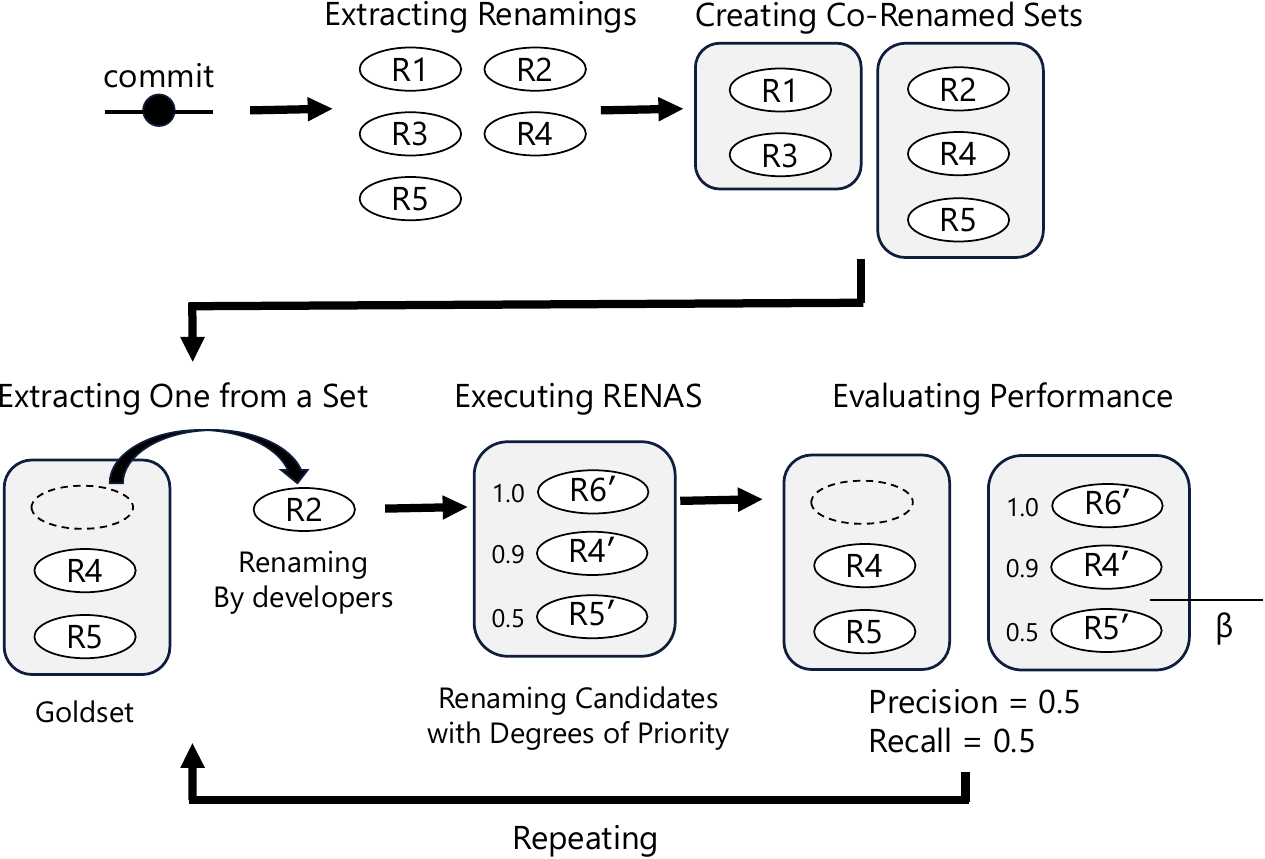}
    \caption{Overview of the evaluation process.}
    \label{fig:abstract-evaluate-technique}
\end{figure}

An overview of the evaluation process is shown in \cref{fig:abstract-evaluate-technique}.
We specifically considered renames within a single commit, so only one developer is involved in the renames.
The dataset of co-renamed sets in \cref{s:evaluationRepository} was utilized to investigate the potential for one renaming in a set to recommend other renamings.

\subsection{Data Collection}\label{s:evaluationRepository}

Two datasets were prepared from GitHub projects as of December 2023: one was automatically identified and the other was manually validated. 

The automatically identified dataset was created by selecting 11 projects from the renaming study conducted by Osumi et al. \cite{Osumi-APSEC2022}.
Co-renamed sets were generated using RefactoringMiner 2.0.2 \cite{Tsantalis-TSE2022}, defined in \cref{s:recommendName} when four or more renamings were obtained in a single commit.
This lower limit was set to emphasize the situations of overlooking co-renamings.
These projects yielded 14,557 renamings and 3,361 co-renamed sets.

\begin{table}[tb]\centering
    \caption{Projects to Validate Manually}\label{tab:visual-repository}
    {\tabcolsep=5pt\begin{tabular}{lccccc}  \hline
       Name & \# Java files & LOC & \# renamings & \# co-renamed sets \\\hline
       \repo{ratpack}  & 831  &  42,160  &  1,424  &  327  \\ 
       \repo{ArgoUML}  &  1,884  &  174,574  & 709  &  178 \\\hline
    \end{tabular}}
\end{table}

The manually validated dataset was developed by selecting two projects from Osumi et al., as outlined in \cref{tab:visual-repository}.
This dataset comprised sets of co-renamed items that were manually validated based on the rename refactorings identified using RefactoringMiner 2.0.2.
Manual validation was conducted by an author.

To create the manually validated dataset, the first step involved checking whether the renamings obtained from the project adhered to the specified relationship.
As described in \cref{s:recommendName}, customary changes, typo corrections, changes in the structure of identifiers, and expansion of abbreviations are excluded from the dataset because they do not follow the relationships.
Only renamings that followed the relationships were included in the co-renamed set.

In cases where an abbreviation was present in the renaming, efforts were made to identify the original words that were abbreviated within the program.
Suppose that the original words could not be located within the program.
In that case, we identified original words by referring to studies on identifier abbreviation \cite{Newman-ICSME2019, Beniamini-ICPC2017}.
Following the expansion of the abbreviation, we determined whether the renaming followed the relationship.

The creation of a manually validated dataset enabled the exclusion of renamings that did not conform to the relationship, which the automatically identified dataset could not exclude.
Additionally, the manual validation process was necessary to address the possibility of inaccuracies in the extracted renaming operations and errors in the formation of co-renamed sets.
There are three examples of renamings that were manually excluded: 
\IdentS{FigSeperator} $\to$ \IdentS{FigSeparator},
\IdentS{GoElementToDependentElement} $\to$ \IdentS{GoElement2DependentElement}, and
\Ident{token} $\to$ \IdentS{theToken}.
The first renaming operation was extracted as \replace([\Ident{sep\underline{e}rator}], [\Ident{sep\underline{a}rator}]) by RENAS.
However, it was determined to be a typo correction and therefore excluded from the dataset. 
RENAS extracted the second renaming operation as \delete([\Ident{to}]).
However, this operation was also excluded as it only altered the structure of the identifier.
The third renaming operation was \Insert([\Ident{the}]).
This operation was excluded from the dataset as it was deemed a customary change based on observations of the renaming of inserting ``the'' before a noun in many commits in the project.

This dataset was utilized to assess the precision of automatically identified results.

\subsection{RQ1: \RQone}\label{s:evaluationRQ1}

\subsubsection{Study Design}\label{ss:evaluationRQ1Method}
The existing approach recommends renamings by tracing the relationships between identifiers until the renaming operation is no longer applicable.
In contrast, RENAS recommends all identifiers for which the relationships can be traced.
We investigated the performance differences resulting from varying recommendation approaches, utilizing precision, recall, and F1-measure as evaluation metrics.

\app{RENAS} was compared with \app{None}, \app{Relation}, and \app{Relation+Normalize}.
\begin{itemize}
    \item \app{None}: Recommending from all identifiers in the source code without considering word inflections and abbreviations.
    \item \app{Relation}: Adding identifiers related to the renamed identifier to the candidates and recommending identifiers to which the renaming operation can be applied from the candidates.
    This method also includes identifiers related to those eligible for renaming.
    Word inflections and abbreviations were not considered.
    This setting was designed to mimic RenameExpander\cite{Liu-TSE2015} for comparison with the proposed approach.\footnote{RENAS cannot be directly compared with RenameExpander owing to their differences in the implementation infrastructure, relationships utilized, and the basic design of the recommendation approach. To facilitate a comparison focusing on the differences in the scoring approach by RENAS while unifying other minor settings, this evaluation setting has been devised.}
    \item \app{Relation+Normalize}: This considers word inflections and abbreviations in the \app{Relation} recommendation approach.
    \item \app{RENAS} (Proposed approach): This approach recommends identifiers with a priority equal to or exceeding the specified threshold.
\end{itemize}

\subsubsection{Results and Discussion}\label{ss:evaluationRQ1Machine}

\begin{figure}[tb]
    \centering
    \includegraphics[width=8cm]{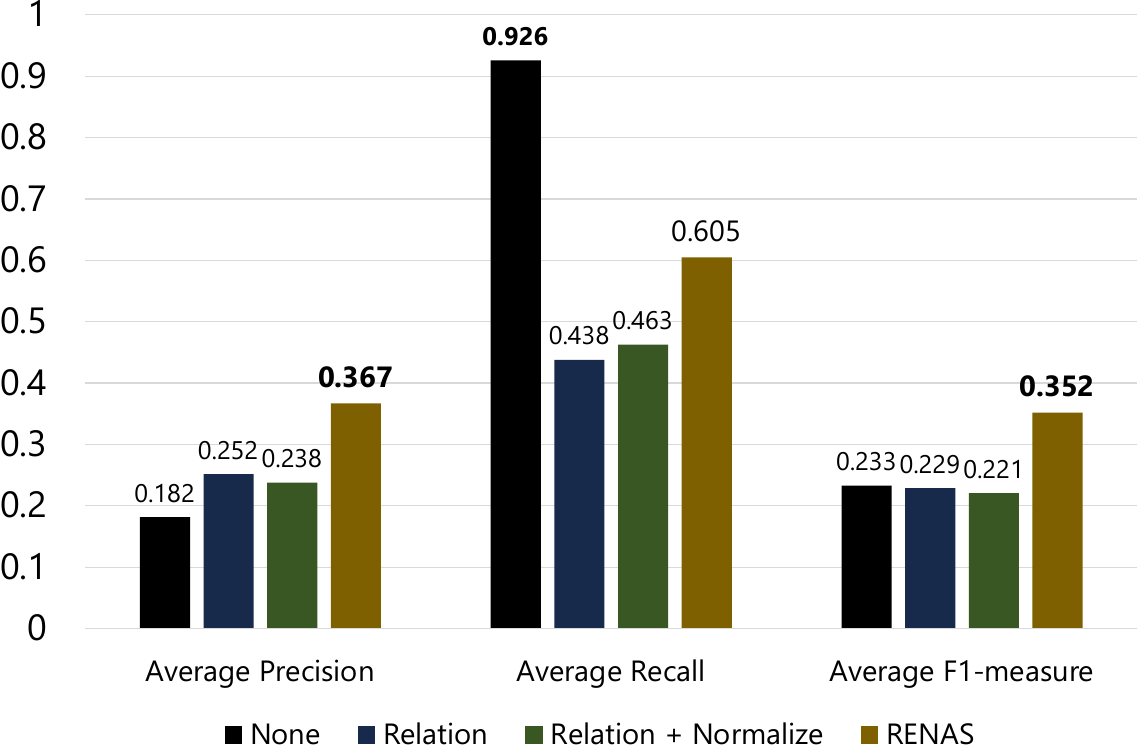}
    \caption{RQ1: Results of the automatically identified dataset.}
    \label{fig:RQ1-automatically-dataset}
\end{figure}

The evaluation results based on the automatically identified dataset are shown in \cref{fig:RQ1-automatically-dataset}.
Each value in this figure represents the average across all projects.

The \app{None} approach achieved the highest recall among the four approaches as they recommend renaming from all identifiers in the source code.
\app{RENAS} achieved the highest precision, and its recall was higher than that of \app{Relation} and \app{Relation+Normalize}.
This can be attributed to the fact that \app{RENAS} considered identifier similarity and suggested renamings for all identifiers that can be traced back to relationships.
For example, identifiers that could not be recommended by tracing the identifiers to which the renaming operation could be applied are shown in \cref{fig:motivation-example-rename2}.
The performance of \app{RENAS} was higher owing to the abundance of such renamings.
Overall, the F1-measure of \app{RENAS} surpassed those of other approaches.

\begin{figure}[tb]
    \centering
    \includegraphics[width=8cm]{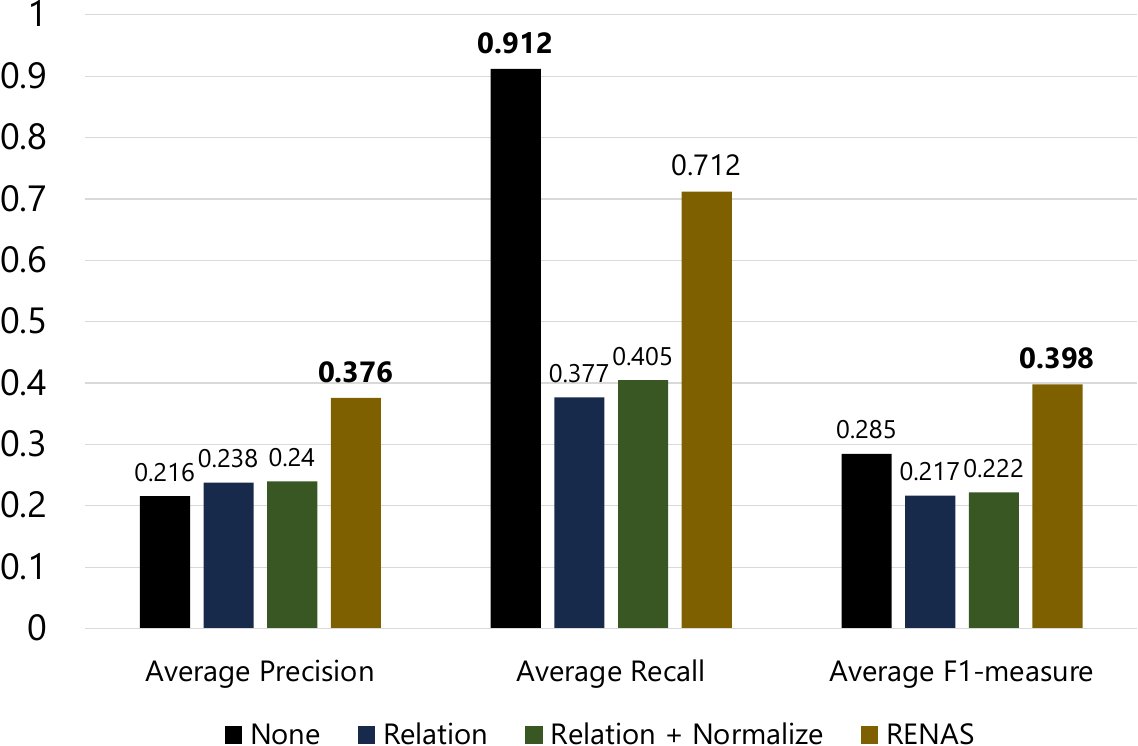}
    \caption{RQ1: Results of the manually validated dataset.}
    \label{fig:RQ1-visual-dataset}
\end{figure}

The results for the manually validated dataset are shown in \cref{fig:RQ1-visual-dataset}.
\app{RENAS} demonstrated the highest precision and F1-measure, whereas \app{None} exhibited the highest recall. 
Therefore, we conclude that \app{RENAS} outperformed the existing approaches.

In our analysis of successful rename recommendations, we distinguished between identifiers that were \emph{directly} traceable (recommended by \app{Relation+Normalize}) and those that were \emph{indirectly} traceable (recommended by \app{RENAS} with $\beta = 0$ but not by \app{Relation+Normalize}).
Consequently, we observed 87,349 directly and 85,007 indirectly traceable identifiers.
These findings underscore the importance of considering indirectly connected identifiers when recommending renamings.

\begin{figure}[tb]\centering
    \begin{minipage}[tb]{0.5\textwidth}\centering
        \includegraphics[width=8.8cm]{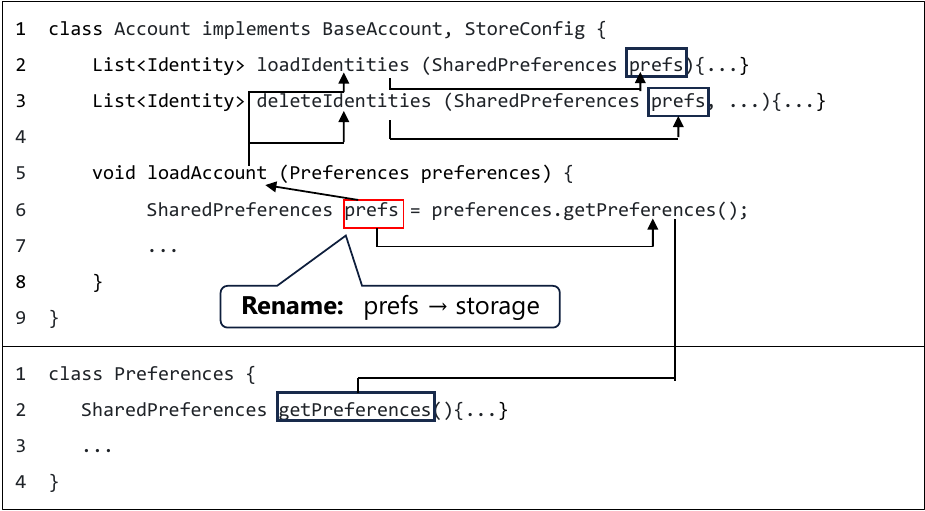}
        \subcaption{Before renaming.}
        \label{fig:rq1_before}
    \end{minipage}
    \begin{minipage}[tb]{0.5\textwidth}\centering
        \includegraphics[width=8.8cm]{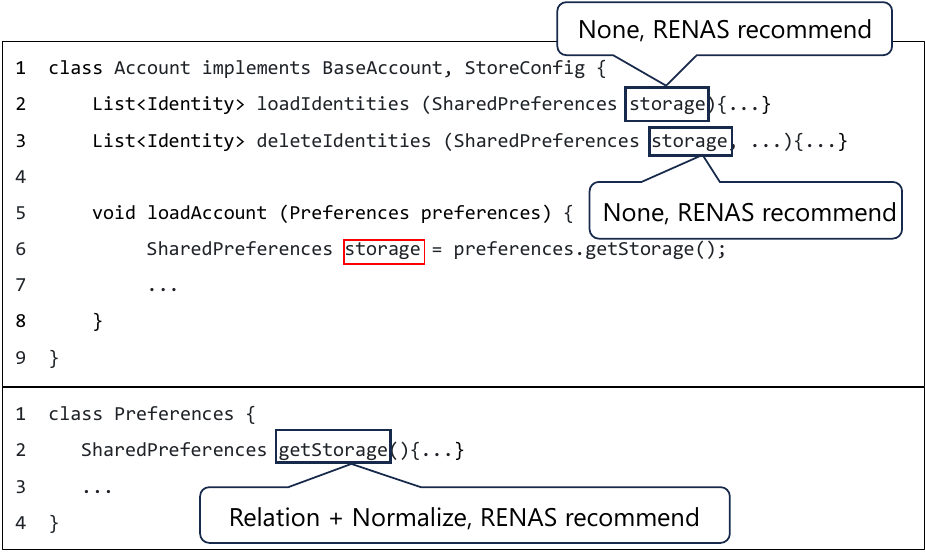}
        \subcaption{After renaming.}
        \label{fig:rq1_after}
    \end{minipage}
    \caption{Recommendation example in \repo{thunderbird-android}.}
    \label{fig:rq1_code}
\end{figure}

An example of successful recommendations for each approach is shown in \cref{fig:rq1_code}.
The source code of \repo{thunderbird-android}\footnote{\url{https://github.com/thunderbird/thunderbird-android}} before and after renaming are shown in \cref{fig:rq1_before,fig:rq1_after}, respectively.
Consider the case in which a developer renames \Ident{prefs} to \Ident{storage} in Line 6.
In this commit, the parameter \Ident{prefs} in Lines 2 and 3 and the method \IdentS{getPreferences} in Line 2 were renamed.
In the case of \app{None}, the parameter \Ident{prefs} in Lines 2 and 3 could be recommended, whereas \IdentS{getPreferences} could not be recommended because identifiers were not normalized. 
Conversely, in the case of \app{Relation}, the extracted renaming operation was \replace([\Ident{prefs}], [\Ident{storage}]) because normalization was not performed.
The identified candidates for renaming were \IdentS{loadAccount} and \IdentS{getPreferences}. 
However, no recommendation was made as neither of them contained \Ident{prefs}.
\IdentS{SharedPreferences} was defined in an external library, which was outside of the scope.
In the case of \app{Relation+Normalize}, the extracted renaming operation was \replace([\Ident{preference}], [\Ident{storage}]) owing to normalization.
This led to the recommendation of renaming \IdentS{getPreferences} to \IdentS{getStorage}. 
In the case of \app{RENAS}, the extracted renaming operation was \replace([\Ident{preference}], [\Ident{storage}]).
The parameter \Ident{prefs} on Lines 2 and 3, as well as the method \IdentS{getPreferences} on Line 2 could be traced through relationships and were recommended because their computed priority exceeded the threshold.

In some cases, \app{Relation+Normalize} succeeded in recommending a renaming, whereas \app{RENAS} failed.
For example, when multiple renaming occurred in a chain, \app{Relation+Normalize} could recommend all the renamings, whereas \app{RENAS} failed owing to low priority when the relationship was traced more than once.
The RENAS method may benefit from adjusting the \cref{eq:distance-priority} of the relationship to enhance its performance.

\Conclusion{%
   \app{RENAS} achieved the highest F1-measure and precision of 0.352 and 0.367, respectively.
   \app{None} reached the highest recall at 0.926.
   We obtained similar results for the manually validated dataset.
   For further performance enhancement, a method for calculating the priority of relationships should be developed.
}

\begin{figure}[tb]
    \centering
    \includegraphics[width=6cm]{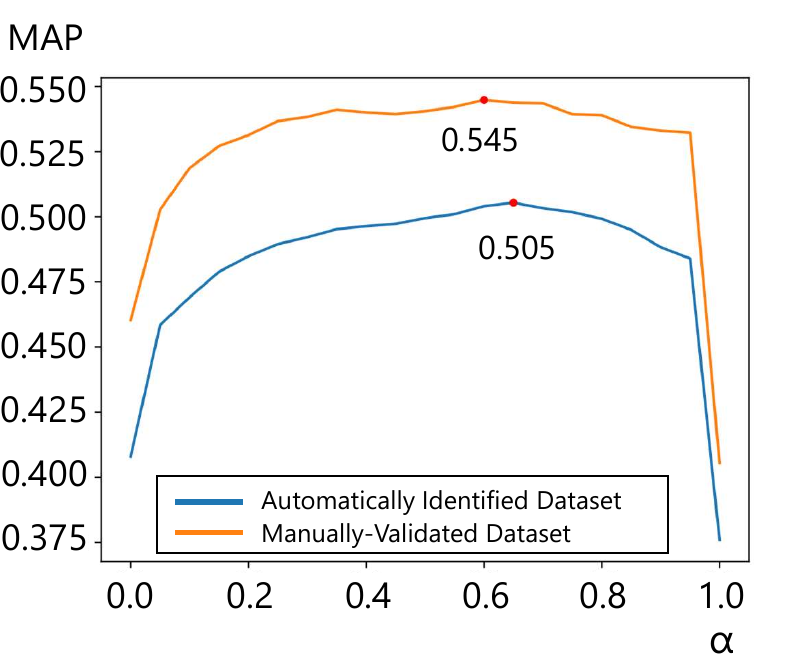}
    \caption{MAP of the automatically identified and manually validated dataset.}
    \label{fig:merge-MAP}
\end{figure}

\subsection{RQ2: \RQtwo}\label{s:evaluationRQ2}
\subsubsection{Study Design}\label{ss:evaluationRQ2method}
We sorted the renaming candidates obtained by running RENAS in order of priority and measured the mean average precision (MAP), mean reciprocal rank (MRR), and Top 1, 5, 10-recall for $\alpha \in \{0,0.05,0.10,\dots,1\}$ of \cref{eq:priority}.
When $\alpha=0$, prioritization is based solely on $\ScoreRel$, whereas when $\alpha=1$, prioritization is determined solely by $\ScoreSim$.
This evaluation aimed to determine the impact of considering both relationships and similarities in prioritization.

In cases where two priorities were equal, we utilized a unique ID as a secondary sorting criterion to establish a complete ranking order for evaluating the effects of prioritization.
The ID is a string comprising the Java file name, identifier name, and type.

\subsubsection{Results and Discussion}\label{ss:evaluationRQ2Machine}

The graph of the MAP is shown in \cref{fig:merge-MAP}, representing the results from the automatically identified dataset and manually validated dataset. 
The horizontal axis of \cref{fig:merge-MAP} represents $\alpha$, whereas the vertical axis represents the MAP.
From the results of the automatically identified dataset, as $\alpha$ increased from 0, the value of MAP also increased.
When $\alpha$ was 0.65, the highest value of MAP was 0.505.
The value of MAP decreased as $\alpha$ increased from 0.65, and at $\alpha=1$, the value of MAP rapidly decreased.
Similar results were obtained for the manually validated dataset.

\begin{figure}[t]
    \centering
    \includegraphics[width=8.5cm]{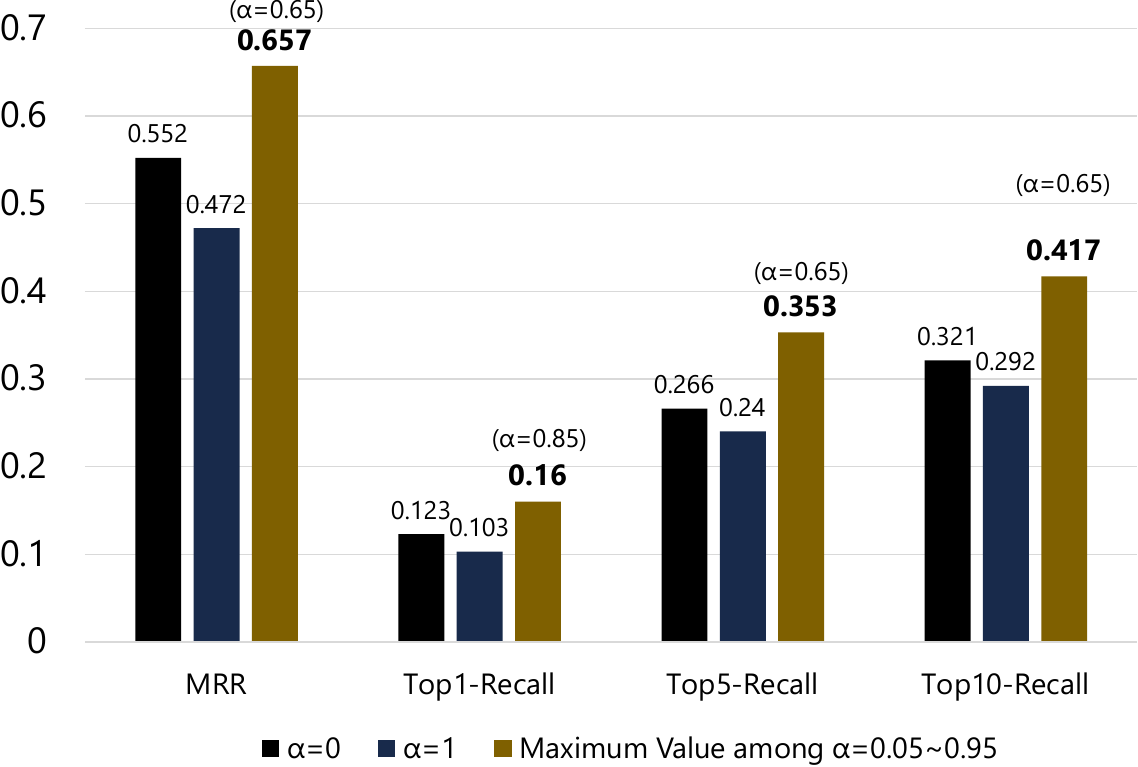}
    \caption{RQ2: Result of the automatically identified dataset.}
    \label{fig:RQ2-automatically-dataset}
\end{figure}

\begin{figure}[t]
    \centering
    \includegraphics[width=8.5cm]{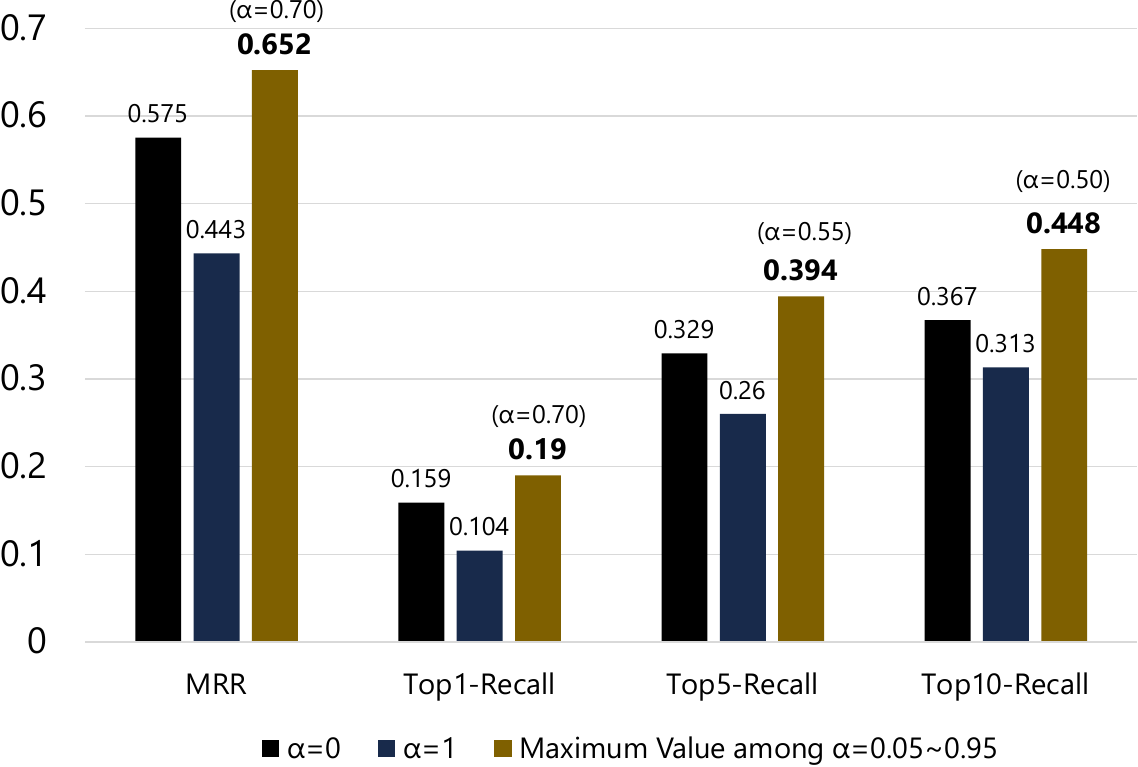}
    \caption{RQ2: Result of the manually validated dataset.}
    \label{fig:RQ2-visual-dataset}
\end{figure}

The values of the other evaluation metrics in the automatically identified and manually validated datasets are shown in \cref{fig:RQ2-automatically-dataset,fig:RQ2-visual-dataset}, respectively.
Although the highest value of $\alpha$ differed, the overall trends in the evaluation metrics remained consistent.

The study revealed that prioritizing recommendations based on both similarity and relationships proved to be more effective across all projects compared with recommendations based solely on similarity or relationships.

RENAS may fail to make successful recommendations for two reasons.
\begin{itemize}
    \item \textit{The recommended identifier has been deleted after the commit.}
    If developers deleted the identifier recommended for renaming after the commit, the recommendation was deemed unsuccessful.
    For example, if the developer renamed \IdentS{ancestorResources} $\to$ \IdentS{matchedResources}, we obtained $\mathit{replace}$([\Ident{ancestor}], [\Ident{matched}]).
    We recommended \IdentS{ancestorURIs} because the identifier existed when our approach traced the relationship.
    However, this renaming recommendation failed because the modification at the commit also deleted \IdentS{ancestorURIs}.

    \item \emph{Missing relationship.}
    Despite our efforts to identify renaming candidates by tracing relationships from the renamed identifier, instances existed in which our approach failed to include certain identifiers in the list of candidates, no matter how extensively we traced the relationships. To recommend them, defining new types of relationships is imperative.
    An example of such a missing relationship is the co-renaming of ``classes in the same package''.
    Introducing such a relationship could potentially enhance recall while reducing precision.
\end{itemize}

\Conclusion{%
    For each evaluation metric, $\alpha \in \{0,1\}$ produced the lowest value, and the highest value was produced within the range of $\alpha \in [0.5, 0.85]$.
    Similar results were obtained for the manually validated dataset.
    The similarity and relationship between identifiers should be considered when calculating the priority.
}

\subsection{Threats to Validity}\label{s:validity}
\subsubsection{Internal Validity}\label{ss:internalValidity}
The performance of RENAS may have been influenced by the accuracy of component tools, such as NLTK, KgExpander, and RefactoringMiner.
For example, RefactoringMiner may generate incorrect renames or lack correct renames, resulting in a decrease in precision.
KgExpander and NLTK may not correctly normalize identifiers.
For example, if a developer renames \Ident{node} to \Ident{entity}, a variable \Ident{n} (as an abbreviation of ``node'') may be the target to be co-renamed.
KgExpander might mistakenly interpret an identifier of \Ident{n} as an abbreviation for ``number'' instead of ``node'', which results in a lack of recommendation.

The automatically identified dataset may not be accurate if the detected renaming operation is incorrect.
To address this concern, we curated a manually validated dataset.
However, a bias may be included in the dataset because the validation was conducted by a single person, potentially resulting in irrelevant renames being included. 
As the dataset is publicly available\cite{dataset,tool}, researchers are encouraged to investigate the impact of this potential threat.

Developers may not have made all necessary renamings in the commits utilized for the evaluation.
Failure to make these renamings could lead to misclassification of true positives to false positives in evaluating renaming recommendations.

\subsubsection{External Validity}\label{ss:externalValidity}
We evaluated RENAS on 13 projects, including manual inspection, and demonstrated that RENAS outperformed existing approaches.
Because the number of projects may be insufficient, we encourage further evaluation on a larger scale to showcase the effectiveness of our approach.

The demonstration of RENAS was conducted only for the Java programming language, and its applicability to other programming languages should be evaluated in future studies.
We believe that RENAS could be applied to other Java-like object-oriented programming languages because similar relationships could be extracted.

\section{Related Work}\label{c:relatedwork}

Numerous studies on the impact of identifier names have been conducted, revealing that names with detailed roles are crucial in understanding programs.
Feitelson et al.\cite{Feitelson-TSE2022} surveyed 334 developers to understand how identifier names are chosen and developed a model for constructing identifier names.
Their findings indicated that identifiers named under similar circumstances varied among developers.
Adhering to the model could lead to the generation of better identifier names.
Hofmeister et al.\cite{Hofmeister-SANER2017} investigated the influence of identifier names with single letters or abbreviations on program comprehension.
They demonstrated that code comprehension was conducted 19\% faster for unabbreviated words.
Beniamini et al.\cite{Beniamini-ICPC2017} investigated the concepts strongly associated with variables named with a single letter.
Avidan et al.\cite{Avidan-ICPC2017} investigated the impact of variable names on program comprehension, highlighting the importance of meaningful names in aiding comprehension.
They noted that parameter names had a more significant influence on program comprehension compared with local variable names.
Some studies have investigated the characteristics of identifiers utilized in large datasets.
Zhang et al.\cite{Zhang-IEEE2020} surveyed 5,000 projects on part-of-speech tags, identifier length, and initialization.
They found that nouns, verbs, and adjectives were frequently utilized in identifiers, with identifier length correlating with importance. 
Additionally, they observed that over 40\% of field and variable names were not initialized at declaration time.
Feitelson et al.\cite{Feitelson-ICPC2023} examined the distribution of the name length and scope of variable identifiers of Java projects.
Their analysis of data from approximately 1,000 Java projects revealed that the wider the scope, the more words incorporated into identifiers, and words comprising six or more letters were often abbreviated.

Developers frequently rename identifiers because they are important, and inappropriate names impair program comprehension.
Arnaoudova et al.\cite{Arnaoudova-TSE2014} demonstrated that renaming identifiers is a complex task, categorizing them into various types based on factors, such as identifier type, change operation, semantic change, and grammatical change. 
They further defined different renaming operations based on this taxonomy.
Peruma et al.\cite{Peruma-IWOR2018} examined 3,795 Java projects to investigate the types of renamings that occurred in class, method, and package names.
Their findings indicated that the meaning of an identifier often became more specific through renaming, and the grammatical structure of identifiers frequently changed.
Moreover, they highlighted that developers' change intentions could be simple or complex.

\begin{table}[tb]\centering
    \caption{Approaches for Recommending Renamings}\label{tab:related-work}
    {\tabcolsep=5pt\begin{tabular}{lccc} \hline
         & Relationships & Abbreviations & Inflection Words\\\hline
       IDD\cite{Deissenbock-SQJ2006}  &   &  &    \\
       Thies et al.\cite{Thief-RSSE2010}  & \checkmark  &  &  \\
       Zhang et al.\cite{Zhang-TOSEM2023} & \checkmark+  & \checkmark & \checkmark \\
       RenameExpander\cite{Liu-TSE2015} & \checkmark++ &  &  \\
       RENAS & \checkmark++ & \checkmark &  \checkmark \\\hline
    \end{tabular}}
\end{table}

The characteristics of the existing approaches and our approach for recommending rename refactorings are listed in \cref{tab:related-work}.
The checkmark (\checkmark) in the table indicates that the element in the column is considered in the approach.
Additionally, the plus sign (+) in the ``Relationships'' column indicates the number of relationship types, with more pluses indicating more relationships.
Deibenbock et al. \cite{Deissenbock-SQJ2006} developed a tool named Identifier Dictionary (IDD) to maintain consistent and concise identifier names.
IDD stores information on all identifiers in the source code and recommends renaming them if it detects identifiers with the same name but different types.
Thief et al.\cite{Thief-RSSE2010} recommended renaming identifiers based on the assignment relationship.
They made recommendations to developers by identifying misspellings, synonyms, and incorrect naming from graphs based on the assignment relationship.
Liu et al.\cite{Liu-TSE2015} established relationships between identifiers and specified candidate identifiers to be renamed based on their relationships with the identifier renamed by a developer.
Among the candidates, they recommended renaming identifiers to which the renaming operation could be applied in the order of the similarity of identifiers.
Zhang et al.\cite{Zhang-TOSEM2023} utilized a score obtained through machine learning derived from identifier relationships, identifier information, and files containing the identifiers.
They also incorporated a score calculated from previous renaming.
RENAS performed recommendation by defining the relationship between identifiers in more detail compared with that of Zhang et al.
However, it does not consider information about files and previous renaming.
Prioritizing identifiers while considering these factors may lead to enhanced accuracy in recommendations. 

Research has been conducted on prioritizing identifier name recommendations.
Kasegn et al. \cite{Kasegn-ICT4DA2021} conducted a study to determine the proximity of words within a document, and when the prefix of an identifier name was provided, their approach recommended the words connected to the prefix in order of their priority.
The recommended identifiers included classes, methods, and fields.
Abebe et al.\cite{Abebe-CSMR2013} analyzed the concepts represented by identifiers, established relationships related to concepts, such as the \emph{isA} and \emph{hasProperty} relationships, and constructed an identifier relationship graph.
Based on this graph, they prioritized their recommendations.
Because RENAS does not rely on concept-based relationships, analyzing these relationships in addition to the current relationships and assigning priority to them may enhance the accuracy.

\section{Conclusion}\label{c:conclusion}
This study proposed an approach named RENAS, which aimed to recommend renaming opportunities by assigning priorities to identifiers based on their similarities and relationships.
The evaluation of RENAS involved comparing its performance with three distinct settings: \app{None}, \app{Relation}, and \app{Relation+Normalize} using an automatically identified dataset and a manually validated dataset.
The results showed that the F1-measure of RENAS was 0.352, which was 0.11 higher than that of the other approaches. 
We observed similar values for the manually validated dataset.
This result suggests that RENAS outperformed existing approaches in terms of effectiveness.
Furthermore, we obtained higher values for each evaluation metric when considering both factors compared with when considering only similarity or relationships.

Several avenues for future research can be explored.
\begin{itemize}
    \item \textit{Creating a Renaming Dataset.}
    In our evaluations, we created co-renamed datasets from commits of projects on GitHub.
    However, developers may not make all the necessary renamings within a commit.
    To evaluate renaming recommendations more accurately, creating an accurate dataset of renamings is necessary.
    \item \textit{Relevance Feedback.}
    Implementing re-prioritization based on developers' adoption of recommended renamings can enhance the accuracy of prioritization.
    \item \textit{Detailed Investigation of Renamings.}
    We will conduct an empirical study to investigate the types of effective co-renaming relationship that impacts the accuracy of co-renaming recommendations and define the relationships based on it.
    Additionally, we will investigate the frequency and reasons for renamings in relation to the difference between direct and indirect relationships.
\end{itemize}

\section*{Acknowledgments}
This study was partly supported by JSPS Grants-in-Aid
for Scientific Research Nos.\ JP24H00692, JP23K24823, JP21H04877, JP21K18302, and JP21KK0179.

\IEEEtriggeratref{13}


\end{document}